\documentclass[letterpaper,english,prd,floatfix,showpacs,twocolumn]{revtex4}
\usepackage{fontenc}
\usepackage[latin1]{inputenc}
\usepackage{float}
\usepackage{graphicx}

\makeatletter

\usepackage{babel}
\makeatother
\begin{document}

\preprint{Rev\TeX{}4 Preprint}

\title{Numerical Solution of the Ekpyrotic Scenario in the Moduli Space
Approximation}

\author{Torquil MacDonald S\o rensen}

\email{t.m.sorensen@fys.uio.no}

\affiliation{Department of Physics, University of Oslo, PO Box 1048, 0316 Oslo,
Norway}

\date{17. Feb 2005}

\begin{abstract}
A numerical solution to the equations of motion for the ekpyrotic
bulk brane scenario in the moduli space approximation is presented.
The visible universe brane has positive tension, and we use a potential
that goes to zero exponentially at large distance, and also goes to
zero at small distance. In the case considered, no bulk brane, visible
brane collision occurs in the solution. This property and the general
behavior of the solution is qualitatively the same when the visible
brane tension is negative, and for many different parameter choices. 
\end{abstract}


\maketitle

\section{Introduction\label{sec:introduction}}

The ekpyrotic scenario was first presented in \cite{Khoury:2001wf}.
Based on string theory \cite{Horava:1995qa,Horava:1996ma}, it is
a cosmological theory containing three-branes in a bulk space with
four spatial dimensions. One of these branes is the observable three-dimensional
universe. The theory was proposed as an alternative to the big bang/inflationary
theory, and avoids the space-time singularity of the big bang. What
is perceived by the inhabitants of the visible three-brane as a big
bang, is explained five-dimensionally as a collision between a bulk
brane and the visible brane. This collision is called ekpyrosis. As
the collision is totally inelastic, this sends the universe into a
hot, expanding phase.

This theory has later been modified into a cyclic universe model,
without a bulk brane \cite{Steinhardt:2001vw,Steinhardt:2001st}.
In the cyclic model, the big bang is realized as a collision between
the visible and the hidden brane. In this scenario, the boundary branes
collide, which means that the collision is accompanied by a strong
space-time singularity, whereby an entire dimension collapses to zero
physical length.

In section \ref{sec:model}, we recall the mathematical basis for
the ekpyrotic scenario. Section \ref{sec:static_solution} contains
the static solution that is the basis for the moduli space approximation.
We describe the moduli space approximation in section \ref{sec:moduli_space_approx}.
The numerical results from solving the moduli space approximation
equations are given in section \ref{sec:numerical_results}. In section
\ref{sec:conclusion} we give our conclusions.

\section{Model\label{sec:model}}

The ekpyrotic scenario is based on Simplified Heterotic M theory,
which is Heterotic M theory with the fields whose equation of motion
allow it, set to zero \cite{Lukas:1998tt,Lukas:1998yy}.

The space-time of the theory is the five-dimensional manifold $\mathcal{M}_{5}\equiv\mathbf{R}^{4}\times\mathbf{S}_{1}/\mathbf{Z}_{2}$,
which we call the bulk space and is given by the following construction.
The $\mathbf{R}^{4}$ part is the usual four-dimensional infinite
space-time, with coordinates $(t,x^{1},x^{2},x^{3})$. The extra dimension
$\mathbf{S}_{1}/\mathbf{Z}_{2}$, with coordinate $y$, is obtained
from the circle $\mathbf{S}_{1}$ as follows. Let $y\in[-R,R]$ with
identification $R\sim-R$ be a coordinate on $\mathbf{S}_{1}$. Let
$\mathbf{Z}_{2}$ be the group of reflections of $\mathbf{S}_{1}$
about $y=0$. We use these actions of $\mathbf{Z}_{2}$ on $\mathbf{S}_{1}$
to define the new manifold $\mathbf{S}_{1}/\mathbf{Z}_{2}$ by identifying
points on $\mathbf{S}_{1}$ that are related by an action of $\mathbf{Z}_{2}$.
Since we then have the identification $y\sim-y$, the coordinate on
$\mathbf{S}_{1}/\mathbf{Z}_{2}$ is $y\in[0,R]$, so this dimension
is a closed line element $[0,R]$. $\mathbf{S}_{1}/\mathbf{Z}_{2}$
is an orbifold, since it consists of the orbits of the group action
of $\mathbf{Z}_{2}$ on $\mathbf{S}_{1}$, and will be referred to
as the orbifold dimension.

Since the extra dimension is a line element, this space-time has two
boundaries, at $y\in\{0,R\}$. Differentiation is not well-defined
on the boundary of a manifold, so we define the action (\ref{eq:het_action})
of the theory on the boundary-less manifold $\mathbf{R}^{4}\times\mathbf{S}_{1}$,
making differentiation well-defined everywhere. We must then demand
$\mathbf{Z}_{2}$ reflection symmetry $y\sim-y$ from all the fields
in the Lagrangian.

The action of the Simplified Heterotic M-theory is\begin{eqnarray}
S & = & \frac{M_{5}^{3}}{2}\int_{\mathcal{M}_{5}}d^{5}x\sqrt{-g}\Big(\mathcal{R}-\frac{1}{2}\partial_{A}\phi\partial^{A}\phi-\frac{3}{2}\frac{1}{5!}e^{2\phi}\mathcal{F}^{2}\Big)\nonumber \\
 &  & -\sum_{i}\frac{3}{2}\alpha_{i}M_{5}^{3}\int_{\mathcal{M}_{4}^{(i)}}d^{4}\xi_{(i)}\Big(\sqrt{-h_{(i)}}e^{-\phi}\nonumber \\
 &  & -\frac{1}{4!}\epsilon^{\alpha\beta\gamma\delta}\mathcal{A}_{ABCD}\partial_{\alpha}X_{(i)}^{A}\partial_{\beta}X_{(i)}^{B}\partial_{\gamma}X_{(i)}^{C}\partial_{\delta}X_{(i)}^{D}\Big),\label{eq:het_action}\end{eqnarray}
where $M_{5}$ is the 5D Planck mass, $g_{AB}$ is the metric on $\mathcal{M}_{5}$,
and $g$ is its determinant. Uppercase Latin indices refer to coordinates
in the bulk space, and small Greek indices refer to coordinates on
a brane. $\mathcal{R}$ is the 5D Ricci scalar, and $e^{\phi}$ is
the size of the Calabi-Yau 6D compact manifold. $\mathcal{A}$ is
a bulk four-form gauge field, with field strength $\mathcal{F}\equiv d\mathcal{A}$.
Furthermore, we have contributions from the three-branes, which couple
to the bulk fields $\phi$ and $\mathcal{A}$. The brane tensions
are $3\alpha_{i}M_{5}^{3}/2$, and their actions are given by the
four dimensional integrals over their world-volumes $\mathcal{M}_{4}^{(i)}$.
$\xi_{(i)}$ are coordinates on the branes. The corresponding integrands
contain the induced metric on their world-volumes, $h_{\alpha\beta}^{(i)}$,
which is the pull-back of the bulk metric onto $\mathcal{M}_{4}^{(i)}$.
$X_{(i)}^{A}(\xi_{(i)}^{\mu})$ are the coordinates in $\mathcal{M}_{5}$
of a point whose coordinates in $\mathcal{M}_{4}^{(i)}$ are $\xi_{(i)}^{\alpha}$.
$\mathcal{A}_{ABCD}\partial_{\alpha}X_{(i)}^{A}\partial_{\beta}X_{(i)}^{B}\partial_{\gamma}X_{(i)}^{C}\partial_{\delta}X_{(i)}^{D}$
is the induced four-form field on the brane world-volume, from the
bulk four-form field $\mathcal{A}$. Anomaly cancellation, which must
be satisfied to enable a consistent quantization of the theory, demands
that the brane tensions sum up to zero \cite{Donagi:2001fs}. Therefore
we parametrize them as $\alpha_{1}=-\alpha$, $\alpha_{2}=\alpha-\beta$
and $\alpha_{3}=\beta$.

\section{Static Solution\label{sec:static_solution}}

A vacuum solution of this theory contains flat, static, parallel branes.
This means that we can express their embeddings as\[
X_{(i)}^{A}=(t,x^{1},x^{2},x^{3},y_{(i)}).\]
We choose brane 1 and 2 as the boundary branes at $y_{(1)}=0$ and
$y_{(2)}=R$. Brane 3 is the bulk brane at $y_{(3)}\equiv Y$. A static
solution of the Euler-Lagrange equations derived from (\ref{eq:het_action})
is \cite{Khoury:2001wf} given by\begin{eqnarray}
ds^{2} & = & D(y)\Big(-N^{2}dt^{2}+A^{2}\sum_{i=1}^{3}(dx^{i})^{2}\Big)\nonumber \\
 &  & +B^{2}D(y)^{4}dy^{2}\label{eq:static_sol_metric}\\
e^{\phi} & = & BD(y)^{3}\label{eq:static_sol_phi}\\
\mathcal{F}_{01234} & = & -NA^{3}B^{-1}D(y)^{-2}D'(y)\label{eq:static_sol_f}\end{eqnarray}
\begin{equation}
D(y)=\Bigg\{\begin{array}{ll}
\alpha y+C & ,y\in[0,Y]\\
\alpha y-\beta(y-Y)+C & ,y\in[Y,R]\\
D(-y) & ,y\in[-R,0]\end{array}\label{eq:d_function}\end{equation}
The moduli space of these solutions is therefore parametrized by the
five constants $N$, $A$, $B$, $C$ and $Y$.

We can obtain the effective scale factor on the visible brane by evaluating
the bulk metric at $y=0$. We obtain\begin{equation}
a_{1}=\sqrt{C}A.\label{eq:four_dim_scale_factor}\end{equation}

From (\ref{eq:static_sol_metric}) we note that the physical length
of the orbifold dimension is given by the expression\begin{eqnarray}
L & = & B\int_{0}^{R}dyD(y)^{2}\nonumber \\
 & = & \frac{1}{3}B\Bigg[\frac{1}{\alpha}\Big((\alpha Y+C)^{3}-C^{3}\Big)+\frac{1}{(\alpha-\beta)}\nonumber \\
 &  & \times\Big(\left((\alpha-\beta)R+C+\beta Y\right)^{3}-(\alpha Y+C)^{3}\Big)\Bigg].\label{eq:length_of_orbifold}\end{eqnarray}

\section{Moduli Space Approximation\label{sec:moduli_space_approx}}

The moduli space approximation involves making a time-dependent solution
by promoting the five constants in the static solution to time-dependent
functions, then substituting this into the original action to obtain
an action for the new time-dependent quantities. The result is only
valid for slowly moving systems, and with insignificant matter production
on the branes.

After integrating over the $y$-direction, the resulting moduli space
approximation Lagrangian is\begin{eqnarray}
\mathcal{L} & = & -\frac{3M_{5}^{3}A^{3}B}{N}\Big[I_{3}\Big(\frac{\dot{A}}{A}\Big)^{2}+\frac{\dot{A}}{A}\Big(I_{3}\frac{\dot{B}}{B}+3I_{2}\dot{C}\nonumber \\
 &  & \hspace{-0.8cm}+3\beta I_{2,b}\dot{Y}\Big)-\frac{1}{12}I_{3}\Big(\frac{\dot{B}}{B}\Big)^{2}+\frac{1}{2}I_{1}\dot{C}^{2}\nonumber \\
 &  & \hspace{-0.8cm}+\beta I_{1,b}\dot{Y}\Big(\dot{C}+\frac{\beta}{2}\dot{Y}\Big)-\beta\Big(\frac{1}{2}D^{2}\dot{Y}^{2}-N^{2}V(Y)\Big)\Big],\label{eq:mod_lagrangian}\end{eqnarray}
where we have added a dimensionless bulk brane potential $V(Y)$ that
was not in the original action (\ref{eq:het_action}). The $I_{m}$
functions are integrals of powers of the function (\ref{eq:d_function})
over the orbifold dimension.\begin{eqnarray}
I_{m,a} & \equiv & 2\int_{0}^{Y}D^{m}dy\nonumber \\
 & = & \frac{2}{\alpha(m+1)}\Big[(\alpha Y+C)^{m+1}-C^{m+1}\Big]\label{eq:i_ma}\\
I_{m,b} & \equiv & 2\int_{Y}^{R}D^{m}dy=\frac{2}{(\alpha-\beta)(m+1)}\times\nonumber \\
 &  & \hspace{-1cm}\Big[((\alpha-\beta)R+\beta Y+C)^{m+1}-(\alpha Y+C)^{m+1}\Big]\label{eq:i_mb}\\
I_{m} & \equiv & I_{m,a}+I_{m,b}\label{eq:i_m}\end{eqnarray}

Since the function $D$ is different on each side of the bulk brane
at $y=Y$, the limit $Y\rightarrow R$ of equation (\ref{eq:i_ma})
is different from the limit $Y\rightarrow0$ of equation (\ref{eq:i_mb}).

The equations of motion are the Euler-Lagrange equations obtained
from the Lagrangian in equation (\ref{eq:mod_lagrangian}). We use
the gauge choice $N(t)=1$ to obtain simpler equations. Note that
since $g_{tt}$ is $-D(y)N^{2}$, cosmic time gauge on the visible
brane at $y=0$ would correspond to the gauge choice $N(t)=1/\sqrt{C(t)}$.

We will solve Euler-Lagrange equations from this action numerically,
and the result will be different from the results in the original
article \cite{Khoury:2001wf}, in which the authors analyzed the equations
after choosing $B(t)$ and $C(t)$ to be constant, because that choice
is not in accordance with the equations of motion.

\section{Bulk Brane Potential}

In \cite{Khoury:2001wf}, it is argued that the bulk brane potential
must satisfy the following conditions:

It must depend only on the $y$ coordinate, since the bulk brane movement
in the orbifold direction must be independent of all the other coordinates
except time. This is so because the brane must remain approximately
flat during its journey across the bulk. Otherwise it would hit the
visible brane at significantly different times on different points
in the visible universe, causing big inhomogeneities in the universe.%
\begin{figure}[H]
\begin{center}\includegraphics[%
  width=8.6cm,
  keepaspectratio]{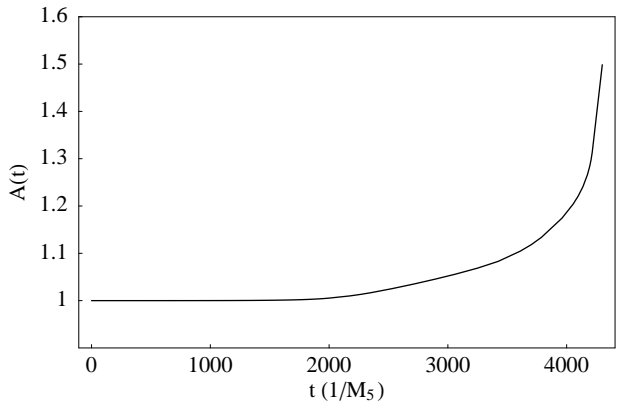}\end{center}

\caption{$A(t)$ increases slowly as the system evolves. \label{fig:a}}
\end{figure}
\begin{figure}[H]
\begin{center}\includegraphics[%
  width=8.6cm,
  keepaspectratio]{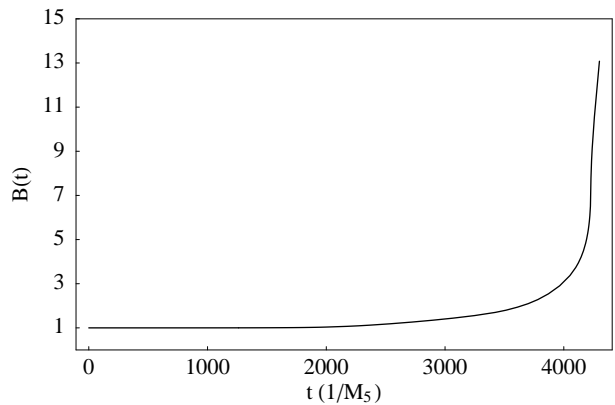}\end{center}

\caption{$B(t)$ starts out slowly, but becomes very steep.\label{fig:b}}
\end{figure}

$V'(y)$ must be small near $y=R$ because otherwise the bulk brane
will not be flat when it is emitted from the hidden brane. The bulk
brane nucleation process starts at one point on the hidden brane,
and grows outwards in all three spatial directions tangent to the
brane. For the bulk brane to become flat, the nucleation growth rate
must be much faster than the movement of the bulk brane in the orbifold
direction.

We must have $V(R)=0$ to prevent inflationary behavior on the hidden
brane.

We need $V(0)=0$ since otherwise it would contribute to a cosmological
constant in the visible universe after the bulk brane collides.

As our dimensionless potential that satisfies the above conditions,
we choose the following.\begin{equation}
V(Y)=-\frac{1}{10}(e^{-5Y/R}-e^{-5})(1-\frac{1}{10Y/R+1})\label{eq:exp_potential}\end{equation}
\begin{figure}[H]
\begin{center}\includegraphics[%
  width=8.6cm,
  keepaspectratio]{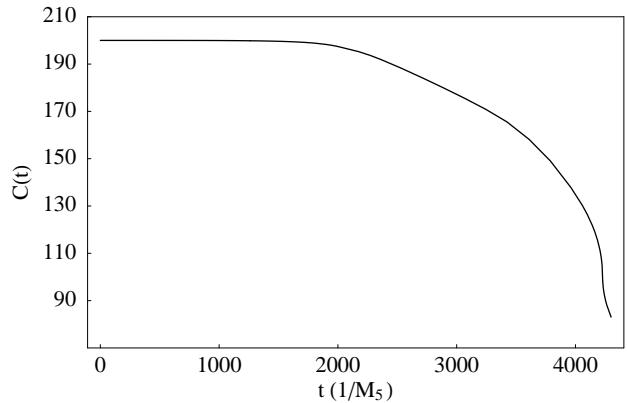}\end{center}

\caption{$C(t)$ falls significantly during the evolution.\label{fig:c}}
\end{figure}
\begin{figure}[H]
\begin{center}\includegraphics[%
  width=8.6cm,
  keepaspectratio]{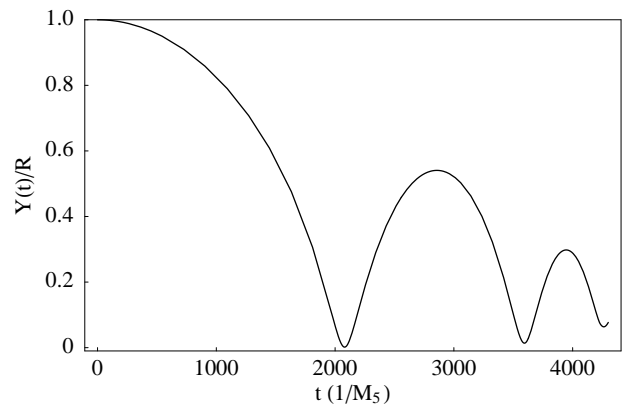}\end{center}

\caption{The bulk brane bounces without colliding.\label{fig:y}}
\end{figure}

\section{Numerical results and discussion\label{sec:numerical_results}}

\begin{figure}[H]
\begin{center}\includegraphics[%
  width=8.6cm,
  keepaspectratio]{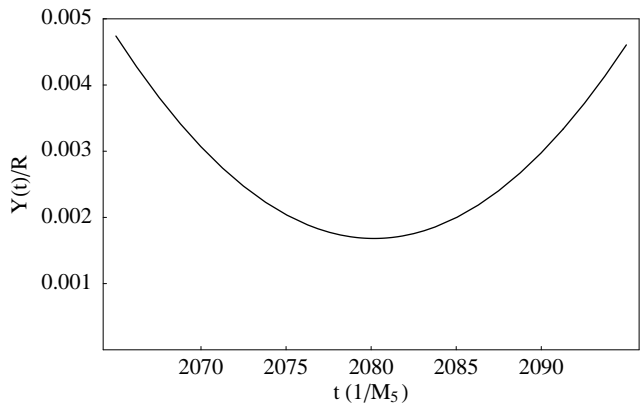}\end{center}

\caption{The bulk brane doesn't collide with the visible universe.\label{fig:y_near}}
\end{figure}

We solve the equations of motion derived from the Lagrangian (\ref{eq:mod_lagrangian}),
numerically from $t=0$ to $t=4300/M_{5}$ in Mathematica\texttrademark,
with parameter values $R=1/M_{5}$, $\alpha=-100M_{5}$, $\beta=M_{5}$.
The initial conditions are $A(0)=B(0)=1$, $C(0)=200$ and $A'(0)=B'(0)=C'(0)=Y'(0)=0$.
We use the potential in equation (\ref{eq:exp_potential}). The solutions
are given graphically in figures \ref{fig:a} - \ref{fig:y_near}.

In the course of the evolution, we see from figures \ref{fig:a} and
\ref{fig:b} that the scale factors $A(t)$ and $B(t)$ increases
monotonically. $C(t)$ in figure \ref{fig:c}, however, decreases
monotonically. These three functions show no oscillatory tendencies,
as opposed to $Y(t)$ in figure \ref{fig:y}, which moves back and
forth in the bulk space. Analysis of the numerical solution shows
that on its first encounter towards the visible brane, the time at
which the separation is the smallest is $t_{c}\approx2080/M_{5}$,
with position $Y(t_{c})\approx0.0017\times R$, as seen in figure
\ref{fig:y_near}. It does not bump into the hidden brane either.
The graph of $Y(t)$ in figure \ref{fig:y} is resembles a damped
cycloidal motion with decreasing wavelength. The amplitude and wavelength
of the oscillations decrease until%
\begin{figure}[H]
\begin{center}\includegraphics[%
  width=8.6cm,
  keepaspectratio]{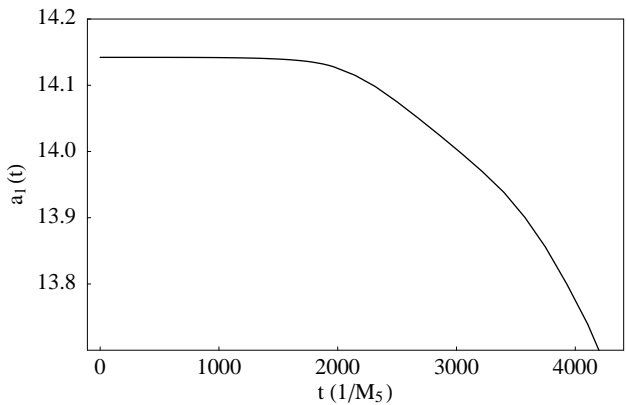}\end{center}

\caption{The four-dimensional effective scale factor decreases as the system
evolves.\label{fig:a_eff}}
\end{figure}
the solution breaks down at around zero wavelength (not shown in the
plot).

The effective scale factor $a_{1}(t)$ of the visible universe, given
by equation (\ref{eq:four_dim_scale_factor}), is plotted in figure
\ref{fig:a_eff}. It decreases as the system evolves.

\begin{figure}[H]
\begin{center}\includegraphics[%
  width=8.6cm,
  keepaspectratio]{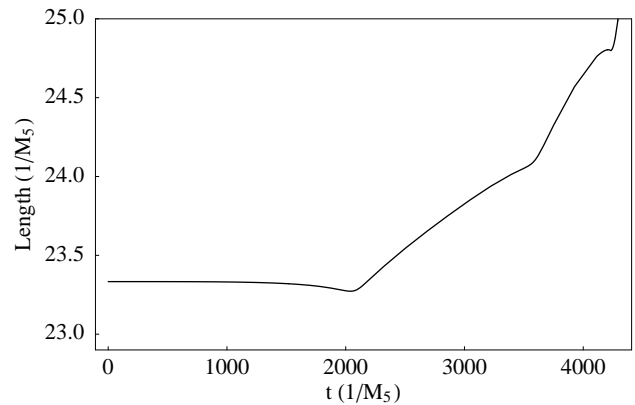}\end{center}

\caption{The physical length of the orbifold dimension increases on the whole
during the evolution, but displays small dips when the bulk brane
is close to the visible brane.\label{fig:length}}
\end{figure}

A plot of the physical length of the orbifold dimension, given by
the expression (\ref{eq:length_of_orbifold}), in figure \ref{fig:length}.
The physical length increases on the whole, but displays small dips
around the times at which the bulk brane is close to the visible brane.

So with these initial values for the unknown functions, there is no
ekpyrosis. Simulations with many other values of the parameters and
initial conditions show similar behavior with no ekpyrosis, so this
is no special case. Using a parabolic potential also gives qualitatively
the same type of solution. A collision can be induced by giving $Y(t)$
a sufficient, small initial velocity, so there no inherent barrier
in the equations at $y=0$ for the bulk brane movement. It is simply
energy conservation that prohibits it from reaching $y=0$.

This suggests that the action (\ref{eq:het_action}) in the moduli
space%
approximation, is not capable of producing a big bang effect, which
is necessary for the ekpyrotic scenario to send the visible universe
into a hot, expanding phase. If there is significant matter production
on the branes during evolution, the moduli space approximation is
inaccurate, but this would probably lead to further energy dissipation
in the bulk brane movement, and therefore not lead to ekpyrosis. For
parameter choices where the branes reach each other within a length
of the order of the Planck length, stringy effect could lead to ekpyrosis
through modifications of the potential or brane capture, even though
they don't meet exactly in the classical solution.

\section{Conclusion\label{sec:conclusion}}

No collision is seen between the bulk brane and the visible brane
in the case considered. The behavior of the solution is qualitatively
the same when the visible brane tension is negative, and also when
the bulk brane potential is parabolic, and also for many other values
of the different parameters. The bulk brane seems to bounce back from
the visible brane, without hitting it, in all cases.

Ways of modifying the model to provoke a collision are e.g. allowing
$\lim_{Y\rightarrow0}V(Y)<0$ , but this conflicts with the conditions
on the potential put forth in \cite{Khoury:2001wf}. A different method
is to introduce extra kinetic energy into the initial condition for
the bulk brane movement. This would introduce an extra fine-tuning
into the theory, which is esthetically undesirable. Ideally, since
the ekpyrotic scenario is built on M theory, the potential should
be calculated from first principles before consequences can be deduced.

\begin{acknowledgments}

I would like to thank my thesis adviser, Prof. \O yvind Gr\o n,
and the Dept. of Physics at the University of Oslo.

\end{acknowledgments}

\bibliographystyle{apsrev}
\bibliography{ekpyrotic_collision}

\end{document}